\pdfoutput=1
\documentclass[sigconf,nonacm]{acmart}
\usepackage{algorithm}
\usepackage{algpseudocode}

\begin{document}

\title{Mixture-of-Experts for Cryptocurrency Order Execution: Training Stability, Tail Risk, and Failure Modes}

\author{Alexander Ardaiz}
\affiliation{%
  \institution{Virginia Tech}
  \city{Blacksburg}
  \state{Virginia}
  \country{USA}}
\affiliation{%
  \institution{Dataism Laboratory for Quantitative Finance}
  \city{Blacksburg}
  \state{Virginia}
  \country{USA}}
\email{ardaiz@vt.edu}

\author{Varun Budati}
\affiliation{%
  \institution{Virginia Tech}
  \city{Blacksburg}
  \state{Virginia}
  \country{USA}}
\affiliation{%
  \institution{Dataism Laboratory for Quantitative Finance}
  \city{Blacksburg}
  \state{Virginia}
  \country{USA}}
\email{varunsb@vt.edu}

\author{Ali Habibnia}
\authornote{Corresponding author.}
\affiliation{%
  \institution{Virginia Tech}
  \city{Blacksburg}
  \state{Virginia}
  \country{USA}}
\affiliation{%
  \institution{Dataism Laboratory for Quantitative Finance}
  \city{Blacksburg}
  \state{Virginia}
  \country{USA}}
\email{habibnia@vt.edu}

\begin{abstract}
Deep reinforcement-learning execution policies vary across training seeds, making architectural gains sensitive to training realisations. We evaluate vanilla Double Deep Q-Learning (DDQL), K-means-partitioned mixtures of DDQL experts (MoE) at $K \in \{2, 4, 8\}$, and dense networks parameter-matched to the $K=4$ and $K=8$ expert budgets on 5-minute mean-aggregated BTC/USDT limit order book data from Binance. No learned configuration significantly improves mean implementation shortfall (IS) over DDQL. Under the reported specification, all have higher mean shortfall than TWAP (0.39 bps) and immediate liquidation (0.21 bps) in frictionless replay with a terminal-urgency penalty that makes early liquidation nearly costless. Complete-run collapse, defined operationally as inactivity before forced liquidation in all 500 evaluation episodes, occurs in 11/100 DDQL runs and none in either MoE $K\geq4$ arm. GPU per-episode inactivity, estimated by an IS-cost-matching proxy without direct action revalidation, rises from 11.1\% (DDQL) to 24.9\% ($K=8$). On a device-matched CPU baseline, annealing reduces DDQL collapse from 12/100 to 0/100 and action-counted episode inactivity to 0/50{,}000. Combining annealed exploration with the aligned reward restores collapse in 19/30 runs; with all three changes, 48/100. Expert partitioning is unnecessary to suppress collapse under the original reward. No MoE $K=8$ run has all 500 evaluation episodes inactive under any of six specifications, yet in five, 1.1--28.7\% of its episodes never trade. Across-seed dispersion is lowest at $K=8$ but non-monotone and not robust to family-wise adjustment, while within-policy tail risk worsens monotonically with $K$. Inference on collapse suppression is sensitive to seed count; significance changes from 30 to 100 seeds do not identify a mechanism.
\end{abstract}



\maketitle

\section{Introduction}
\label{sec:intro}
Execution costs separate an investment strategy's hypothetical returns from its realised ones: commissions, limited book liquidity and reactive pricing open a gap that Perold \cite{perold1988} formalised as the \emph{implementation shortfall}. It grows with order size, since a large market order walks progressively deeper book levels while the interest it reveals invites adverse quote adjustments, making execution cost a first-order determinant of realised performance for institutional investors. The classical response schedules a parent order into child orders over time, equally sized (TWAP) or volume-proportional (VWAP), refined by the optimal execution literature \cite{bertsimas1998, almgren2001} and, given the problem's dynamic-programming structure and high-dimensional book state, by reinforcement learning (RL) from Nevmyvaka et al.\ \cite{nevmyvaka2006} through modern deep Q-learning variants \cite{ning2021, macri2024}.

Most of this literature was developed for public equity markets. Cryptocurrency markets differ along dimensions that matter for execution: continuous trading without auctions, fragmented and comparatively thin liquidity, retail-dominated flow, and rapidly shifting volatility and spread regimes \cite{almeida2024, easley2024, westland2021}, creating a need for execution research evaluated on cryptocurrency LOB data \cite{schnaubelt2022}.

A further challenge, and the one this paper is organised around, is optimisation instability. Deep reinforcement-learning execution policies can vary substantially across training seeds even when architecture, data and evaluation windows are held fixed, so a reported architectural advantage may reflect a favourable training realisation rather than a reproducible property of the architecture. The Mixture of Experts (MoE) paradigm \cite{jacobs1991, shazeer2017} maintains several policies behind a router that dispatches each situation to one of them; Li et al.\ \cite{li2024moe} recently demonstrated its promise for neural order execution. We treat it here as a candidate mechanism for distributing learning across experts rather than as a regime classifier, and combine an unsupervised K-Means episode partition with a bank of Double Deep Q-Learning (DDQL) experts, evaluated on a temporally disjoint out-of-sample window of BTC/USDT limit order book data mean-aggregated to five-minute bars, repeated over 100 independently trained runs to quantify seed-level variance.

The contributions are fourfold. First, across 100 independently trained runs evaluated on a common set of out-of-sample windows, no learned configuration significantly improves mean implementation shortfall over DDQL: at $K{=}8$ the estimated mean difference is $-0.03$ bps, 95\% CI $[-0.15, +0.09]$, and the data do not establish equivalence. Second, we examine the apparent effect of expert partitioning on collapse under the reported training specification. Because episodes start at independently sampled bars, the partition is uninformative about the routed episode's conditions at inference, so regime conditioning cannot be the mechanism; 11 of 100 vanilla-DDQL runs meet the complete-run-collapse criterion, whereas no MoE run at either $K{=}4$ or $K{=}8$ does (Fisher $p{=}0.0007$ for each). Complete-run collapse is defined operationally as inactivity before forced liquidation in all 500 evaluation episodes, without establishing that every expert collapses. In this GPU campaign, the IS-cost-matching proxy estimates that per-episode inactivity rises with $K$, from 11.1\% for DDQL to 24.9\% at $K{=}8$, and 84 of 100 $K{=}8$ runs contain never-trade episodes. Decomposing the training specification on a device-matched CPU baseline shows that annealed exploration alone reduces DDQL's complete-run collapse from 12/100 to 0/100 (exact McNemar $p{=}4.9{\times}10^{-4}$) and its episode-level never-trading to none of 50{,}000 evaluation episodes, counted from executed quantities, demonstrating that expert partitioning is not necessary to suppress this failure under the as-reported specification. Third, the specification changes interact non-additively: annealed exploration alone removes complete-run collapse, but adding the aligned reward restores it in 19 of 30 runs, rising to 48 of 100 when all three changes are applied together, while changing replay retention produces no detectable effect on collapse rates. No MoE $K{=}8$ run has all 500 evaluation episodes inactive under any of the six specifications, but in five of them 1.1--28.7\% of its episodes never trade (CPU campaigns), so this pattern too reflects counting by run. Across-seed dispersion is lowest at $K{=}8$ but non-monotone and not robust to family-wise adjustment, and episode-level tail risk moves the other way, worsening monotonically with $K$ (per-run CVaR-95 at $K{=}8$: $+11.5$ bps versus DDQL, 95\% CI $[+3.7, +19.7]$). Fourth, we report in full the negative results and confounds found along the way: a corrected router-lag bug, a fill-price floor that never binds, a TWAP reference beating every learned configuration under the specification as reported, and inference on collapse suppression that is sensitive to the difference between 30 and 100 seeds.

Section~\ref{sec:lob} covers the limit order book and simulation; Section~\ref{sec:oe} reviews order execution and the MoE approach; Section~\ref{sec:design} describes the experimental design, models, and data; Section~\ref{sec:results} presents results; Section~\ref{sec:conclusion} concludes with limitations and future work.

\section{Limit Order Book}
\label{sec:lob}

Limit orders rest at a stated price and size, ranked by price then time priority, and the gap between best bid and best ask is the spread. Market orders cross the spread and consume progressively deeper resting liquidity, generating spread- and depth-related execution costs; in live markets the revealed interest additionally induces transient and persistent price impact as other participants adjust their quotes \cite{cartea2015, bouchaud2018}, a feedback channel our replay does not model (Section~\ref{sec:data}). Purpose-built simulators are central to this literature because latency, fill dynamics and permanent impact are costly to reproduce faithfully. A common simplification is that small orders leave negligible permanent impact. Reactive agent configurations of ABIDES \cite{byrd2020abides} allow background order flow to respond to an execution agent \cite{karpe2020,vyetrenko2020}. JAX-LOB \cite{frey2023jaxlob} is a GPU-accelerated matching engine that can process historical or strategic-agent messages; historical replay alone captures direct book effects but leaves background order flow unresponsive to the agent. Our environment (Section~\ref{sec:data}) replays historical book data directly rather than synthesising a background population, a choice we return to in Section~\ref{sec:limitations}.

\section{Order Execution}
\label{sec:oe}
\subsection{Treatment of Order Execution}
\label{sec:oe_treatment}
Bertsimas and Lo \cite{bertsimas1998} founded the modern optimal execution literature via stochastic dynamic programming, showing naive equal-sized liquidation is optimal only under linear permanent impact and random-walk prices; Almgren and Chriss \cite{almgren2001} extended this to a mean--variance efficient frontier parallel to Markowitz \cite{markowitz1952}. The dynamic-programming structure and high-dimensional book state made execution a natural reinforcement-learning application \cite{suttonbarto2018, rao2022}: Nevmyvaka et al.\ \cite{nevmyvaka2006} first beat benchmark schedules with Q-learning on equity LOB data, and Hendricks and Wilcox \cite{hendricks2014} learned deviations from an Almgren--Chriss base policy. Deep Q-learning \cite{mnih2015} and its double-network variant \cite{vanhasselt2016, ning2021} replaced the tabular Q-function; Macr\`{\i} and Lillo \cite{macri2024} paired Double DQN with liquidity-varying impact, and recent work extends these ideas to cryptocurrency markets and mixed action spaces \cite{schnaubelt2022, cheridito2025}.

\subsection{Mixture of Experts Approach}
\label{sec:oe_moe}
We evaluate a Mixture of Experts (MoE) architecture \cite{jacobs1991, shazeer2017} based on the design suggested by Li et al.\ \cite{li2024moe}. Rather than training a single agent to handle all market conditions, the system maintains $K$ partitioned expert policies and routes each execution episode to one of them according to a cluster assignment from the routing heuristic.

The MoE architecture consists of two primary components:
\begin{enumerate}
  \item \textbf{Expert Q-Networks} ($E_1, \ldots, E_K$): A set of $K$ independent DDQL agents, each with the same 5-layer fully connected architecture as the vanilla DDQL model. Each expert is trained exclusively on the episodes assigned to it by the K-Means partition. As established below, at inference this partition is uninformative about the routed episode's conditions; we retain the architecture as specified by Li et al.\ \cite{li2024moe} and test empirically what the partition delivers.
  \item \textbf{K-Means Router}: A K-Means clustering model fitted on episode-level market features ($K{=}4$ in the default configuration; we vary $K \in \{2, 4, 8\}$ in the ablation of Section~\ref{sec:results}). At inference time, features from the \emph{previously processed} episode are mapped to the nearest cluster centre. Episode order is computational, not chronological, because episode starts are sampled independently. The design intent is to identify the prevailing regime without look-ahead; we show below that it cannot. The corresponding expert then selects the action for the current episode. A soft-gating variant, averaging expert Q-values by inverse-distance weights, is supported but not evaluated here; all experiments use hard gating.
\end{enumerate}

Let $\mathbf{f}_{e-1}$ denote the preceding episode's standardised feature vector, $\boldsymbol{\mu}_k$ cluster $k$'s centroid, $s_t$ the current state and $q_t$ remaining inventory. Under hard gating, the assigned expert and its greedy action are:
\begin{equation*}
k^*_e = \arg\min_k \| \mathbf{f}_{e-1} - \boldsymbol{\mu}_k \|_2, \qquad a_t = \arg\max_{v \in \{0,\ldots,q_t\}} Q_{k^*_e}(s_t, v).
\end{equation*}

Because each episode's starting bar is sampled independently of the preceding one (Section~\ref{sec:design}), the lagged routing features are uninformative about the current episode's conditions by construction, in training and evaluation alike. The router therefore partitions episodes without directly conditioning on the routed episode's regime. Its lagged features can nevertheless depend on the preceding policy's stopping time (Section~\ref{sec:moe_model}). Section~\ref{sec:phase2} bounds direct routing changes at evaluation with trained experts fixed; effects through training remain unmeasured.

\section{Experimental Design}
\label{sec:design}
We adopt the expert-training principle of Li et al.\ \cite{li2024moe}, a router dispatching each episode to a specialized expert, but our execution protocol, routing mechanism and evaluation are our own. Each model's execution performance is measured by implementation shortfall (IS) relative to the arrival mid-price and compared with a time-weighted average price (TWAP) schedule.

Each episode liquidates $q_0{=}20$ source-data quantity units over at most $N{=}10$ decision steps, indexed $t=0,\ldots,N{-}1$, ending early if inventory reaches zero. The source quantity unit has not been independently verified. The order is large relative to displayed liquidity: median visible ten-level depth is 20.9 units against the 20-unit parent order, so sufficiently aggressive child orders can approach or exceed the displayed depth (Section~\ref{sec:phase2}). Each step is one 5-minute bar (a $\sim$50-minute episode), the starting bar sampled uniformly at random from the relevant window, and at each step the agent chooses how many remaining units to sell via market order. All chosen sales are market orders. In all but the final two steps, the order walks the book from the best bid; in the final two steps, an imposed terminal-urgency rule starts the walk at the fifth-best bid. Any inventory remaining at the horizon is force-liquidated. Section~\ref{sec:moe_model} (Actions) gives the fill mechanics precisely as pseudocode.

Our contribution is not a new routing architecture but a repeated-training evaluation separating seed sensitivity, capacity, schedule-baseline performance and tail risk. Li et al.\ \cite{li2024moe} mix neural execution policies by differentiable soft gating in an equity environment, asking how capacity scales; Xu et al.\ \cite{xu2025moedpo} mix four deterministic broker strategies within ABIDES-Gym, integrated by preference optimisation, asking how execution preferences can be made interpretable. We partition learned DDQL networks by an unsupervised clustering router with isolated buffers on BTC/USDT LOB replay, asking what repeated seeds reveal about training stability and tail risk. Their use of ABIDES-Gym indicates the higher-fidelity replication we defer to future work is tractable.

\subsection{MoE Model}
\label{sec:moe_model}
The Mixture of Experts model is trained in two stages:

\begin{description}
    \item[Warm-up stage:] The agent executes 300 TWAP episodes without learning. The experiment driver extracts realised volatility from observed post-step mid-prices, mean quoted spread divided by their mean, mean book imbalance over the preloaded window, and a fourth coordinate fixed at 1.0. Standardisation maps the constant coordinate to zero, leaving three effective dimensions (Section~\ref{sec:phase2}). During learning and evaluation, selling out early truncates the observed price path; volatility and the spread denominator therefore depend on the policy's stopping time. On these warm-up features, K-Means ($K{=}4$, k-means++ initialisation~\cite{arthur2007}) is fitted on the $z$-score--normalised feature matrix. Clustering functions here as a routing heuristic rather than a claim of cleanly separated regimes: cryptocurrency LOBs exhibit overlapping and rapidly shifting dynamic states \cite{li2023entropy}, and silhouette values~\cite{rousseeuw1987} on the warm-up set are correspondingly modest.

    \item[Specialisation stage:] Each episode is assigned to the centre nearest the \emph{previous} episode's feature vector, the same rule used at inference, so training and evaluation share one routing mechanism. Transitions are stored only in the corresponding expert's replay buffer, and standard DDQL updates (Double Q-learning with target network, MSE loss) are applied per expert. Each expert therefore trains on a disjoint subset of episodes, which Section~\ref{sec:oe_moe} shows is not regime-homogeneous with respect to the episodes it is subsequently used to execute. The dense training loop halves its 15{,}000-transition replay buffer when full; the MoE loop never invokes that operation. Training supplies at most 25{,}000 transitions, with the actual count depending on episode stopping times. The arms therefore differ in replay retention as well as in partitioning.
\end{description}

\textbf{Architecture:}
\begin{itemize}
    \item \textbf{Experts and router:} $K$ independent Q-networks ($K{=}4$ by default), each a 5-layer fully connected network with 30 hidden nodes and LeakyReLU activations, identical to the vanilla DDQL architecture. Routing requires no neural network: assignment is by nearest centroid in the standardised feature space, effectively three-dimensional in this experiment.
    \item \textbf{State:} The state is 4-dimensional: normalised inventory $\tilde{q}_t$, normalised time $\tilde{t}$, normalised mid-price $\tilde{S}_t$, and the Bid-Ask (BA) spread fraction. The BA spread lets the agent react to immediate liquidity constraints. The representation excludes depth-of-book sizes; the policy observes only these four state variables.
    \item \textbf{Actions:} Discrete $v_t \in \{0, 1, \ldots, q_t\}$ units to sell at step $t$, filled by walking the 10-level depth-of-book record for that bar. Algorithm~\ref{alg:fill} gives the fill rule exactly, including the two places a deep-book extrapolation can bind: mid-episode overflow beyond level 10 (line~11), and terminal liquidation of any remainder at the horizon (line~17). Both are floored at a fraction $\phi$ of the prevailing reference price, $\phi{=}0.5$ by default; Section~\ref{sec:floor} shows the floor never binds at our order size.

\begin{algorithm}[t]
\caption{Market-order fill and inventory update at decision step $t$}
\label{alg:fill}
\begin{algorithmic}[1]
\Require inventory $q_t$, chosen quantity $v_t \in \{0,\ldots,q_t\}$
\State $\textit{start} \gets 4$ if $t \geq N{-}2$ (terminal-urgency phase) else $0$
\State $\textit{remaining} \gets v_t$, $\textit{cash} \gets 0$
\For{$\ell = \textit{start}, \ldots, 9$} \Comment{walk visible bid levels}
  \State $\textit{take} \gets \min(\textit{remaining}, \textit{size}_\ell)$
  \State $\textit{cash} \gets \textit{cash} + \textit{take} \cdot \textit{price}_\ell$
  \State $\textit{remaining} \gets \textit{remaining} - \textit{take}$
  \If{$\textit{remaining} = 0$} \textbf{break} \EndIf
\EndFor
\If{$\textit{remaining} > 0$} \Comment{excess beyond visible depth only}
  \State $p_{\text{floor}} \gets \max(\textit{price}_9 - 5\,\Delta,\ \phi \cdot S_t)$
  \State $\textit{cash} \gets \textit{cash} + \textit{remaining} \cdot p_{\text{floor}}$
\EndIf
\State $q_{t+1} \gets q_t - v_t$ \Comment{inventory updates once per step}
\If{$t = N{-}1$ \textbf{or} $q_{t+1} = 0$} \Comment{episode ends}
  \If{$q_{t+1} > 0$} \Comment{forced liquidation of what is left}
    \State $p_{\text{floor}} \gets \max(\textit{price}_9 - 10\,\Delta,\ \phi \cdot S_{t+1})$
    \State $\textit{cash} \gets \textit{cash} + q_{t+1} \cdot p_{\text{floor}}$
    \State $q_{t+1} \gets 0$
  \EndIf
\EndIf
\end{algorithmic}
\end{algorithm}

\noindent Here $\Delta = (\textit{price}_0 - \textit{price}_9)/9$, used at lines 11 and 17, is the average per-level price step observed in that bar's book. Two properties bear on results reported later. First, the action set is $\{0,\ldots,q_t\}$ at \emph{every} step including $t=N{-}1$: complete liquidation is enforced by forced terminal liquidation (lines 16--19), not by restricting the agent's choice, which is what makes a never-trade policy feasible and gives it a well-defined cost. Second, the fifth-best-bid start in the final two steps is not a market-order fill in the usual sense but an imposed late-liquidation penalty that discards better available quotes by construction, favouring earlier execution mechanically and independently of market conditions. We verified this pseudocode numerically: across all 500 evaluation episodes and six probe policies, realised cash matches an independent implementation to machine precision, voluntary sales plus forced liquidation always equal $q_0$, and inventory is never negative.
\end{itemize}

\subsection{DDQL Model}
\label{sec:ddql_model}
The Double Deep Q-Learning (DDQL) agent follows the architecture proposed by Macr\`{\i} and Lillo \cite{macri2024}, adapted for our optimal execution environment:

\begin{description}
    \item[Algorithm:] We use the Double DQN update to reduce maximisation bias, minimising mean squared error between current and target Q-values. Let $r_t$ be the immediate reward, $\gamma$ the discount factor, and $d_t$ indicate an episode-ending transition. The online network selects the next action and the target network evaluates it:
    \[
    y_t = r_t + \gamma(1-d_t) Q_{\text{target}}(s_{t+1}, \arg\max_a Q_{\text{main}}(s_{t+1}, a))
    \]
    \item[Exploration Strategy:] We use an $\epsilon$-greedy policy in which exploratory actions are drawn from a Binomial distribution $B(q_t, 1/(N{-}t))$. The expected exploratory sale is $q_t/(N{-}t)$, the remaining inventory divided by the remaining steps.
    \item[Experience Replay:] We implement a memory buffer with a ``halving'' mechanism. When the buffer reaches capacity, the oldest half of the transitions are discarded to maintain relevance to the current policy.
\end{description}

\begin{table}[t]
\centering
\caption{Full training configuration.}
\label{tab:hyperparams}
\footnotesize
\begin{tabular}{ll}
\toprule
\textbf{Parameter} & \textbf{Value} \\
\midrule
Network & 5 layers, 30 hidden, LeakyReLU \\
Optimizer & Adam, lr $=10^{-4}$ \\
Batch / buffer & 32 / 15{,}000 (halved on overflow) \\
Target sync & every 100 gradient updates \\
Discount $\gamma$ & 1.0 (risk-neutral) \\
$\epsilon$ schedule & $\times 0.9995$ per 100 actions, floor 0.01 \\
 & (after 25{,}000 actions: $\epsilon\approx0.88$) \\
Episodes (warmup/train/eval) & 300 / 2{,}500 / 500 \\
Eval seeds & $1000+e$, shared across models \\
Training seed & $\text{run}\times 42$, shared DDQL/MoE \\
K-Means & $k$-means++ init, seed 42 \\
\bottomrule
\end{tabular}
\end{table}

\subsection{Cryptocurrency Data}
\label{sec:data}
We evaluate on cryptocurrency limit order book data:

\begin{description}
    \item[Source:] Bitcoin (BTC/\allowbreak USDT) limit order book data originating from Binance and sourced from Kaggle, spanning January 9--20, 2023, with 10 levels of depth on each side of the LOB. The training window (January 9--14) is used to fit the agents and seed the cluster centroids, while a temporally disjoint evaluation window (January 16--20) is strictly reserved for out-of-sample testing; the intervening period is held out so that no information leaks across the train/test boundary.
    \item[Preprocessing:] The raw snapshot data is mean-aggregated into 5-minute bars, yielding mid-price, bid-ask spread, per-level bid prices and sizes, and a volume-imbalance series.
    \item[Simulation:] This is \emph{not} a calibrated Almgren--Chriss simulator: each bar's mid-price and 10-level bid book are replayed directly from the data (Algorithm~\ref{alg:fill}); no random-walk price process and no calibrated $\kappa_t,\alpha_t$ enter the evaluation at any point. Every column, including each level's price and size, is aggregated by its 5-minute mean, so a bar's book is a time-averaged book rather than an instantaneous one. The replay is frictionless: the agent's trades have no feedback on later prices or books, a deliberate midpoint between a synthetic simulator and full agent-based replay (Section~\ref{sec:lob}).
\end{description}

\section{Experimental Results}
\label{sec:results}

To evaluate the efficacy of the proposed \textbf{Mixture of Experts (MoE)} and \textbf{Double Deep Q-Learning (DDQL)} models, we benchmark their performance against industry-standard execution strategies. Section~\ref{sec:phase2} is the out-of-sample evaluation on the temporally held-out January 16--20 period, which carries the paper's central claim; Section~\ref{sec:decomp} tests which specification components and interactions are implicated in the failure exposed by that evaluation.

\subsection{Baseline Strategies}

We use two model-free references. \textbf{TWAP} liquidates the order $Q$ in equal batches $q_t = Q/T$ at each interval, and \textbf{immediate liquidation} sells the full inventory at the first step. VWAP is not evaluable here: the dataset carries no traded-volume series from which to build a volume profile.

\subsection{Performance Metric: Implementation Shortfall}

The core metric is \textbf{Implementation Shortfall (IS)} \cite{perold1988}. Let $S_0$ be the arrival mid-price, $Q$ the initial quantity, and $C$ total execution proceeds, including any forced liquidation at the horizon. Then IS in basis points is:
\[ \text{IS} = 10^4\frac{S_0Q-C}{S_0Q}. \]
Positive IS is a cost; lower is better.

\subsection{Out-of-Sample Evaluation}
\label{sec:phase2}
Agents train on the January 9--14 period and are evaluated exclusively on the temporally disjoint January 16--20 period (Section~\ref{sec:data}). To separate genuine architectural effects from seed-level noise, we repeat the full train-and-evaluate pipeline for 100 independently trained runs (training seed $=\text{run}\times 42$), holding the same 500 evaluation episodes fixed across every run and every model (eval seed $=1000+e$). An evaluation episode is a deterministic function of its seed alone, so this is guaranteed by construction rather than by convention, and it means identical evaluation episodes across models, enabling paired comparisons of realised execution costs on the same market windows (Table~\ref{tab:hyperparams}). Runs are also matched by training-seed index, but architectures consume random draws differently, so we report unpaired tests alongside paired ones. Because every policy is evaluated on the same market windows, the across-run uncertainty below reflects training-seed variability conditional on this evaluation sample, not sampling uncertainty over alternative market periods.

\textbf{Capacity-matched control.} Since $K$ experts carry $K\times$ the parameters, we also train single DDQL networks matched to the \emph{total} $K{=}4$ and $K{=}8$ expert budgets (widths 64 and 93, within 5\% of target), denoted DDQL-cap4 and cap8.

\begin{table}[t]
\centering
\caption{100-run out-of-sample ablation (Implementation Shortfall, bps). DDQL-cap$k$ is a single network parameter-matched to MoE $K{=}k$'s total capacity. ``SD of run-level mean IS'' is the across-seed dispersion of the 100 per-run means; CVaR-95 is the mean IS at or above each run's empirical 95th percentile (ties included), computed over its 500 episodes and then averaged across runs. Complete-run collapse denotes inactivity before forced liquidation in all 500 evaluation episodes; classification uses an IS vector identical to the never-trade reference (see text). TWAP and immediate liquidation are deterministic given the 500 windows, so they have no run-level dispersion.}
\label{tab:ablation}
\scriptsize
\setlength{\tabcolsep}{3.5pt}
\begin{tabular}{l c c c c c}
\toprule
\textbf{Model} & \textbf{Mean} & \textbf{Paired diff vs DDQL} & \textbf{SD of run-} & \textbf{CVaR-} & \textbf{Collap-} \\
               & \textbf{IS}   & \textbf{(95\% CI)}            & \textbf{level mean IS} & \textbf{95} & \textbf{sed} \\
\midrule
Immediate liq.\ & 0.21 & ,  & , & 0.8 & , \\
TWAP            & 0.39 &, & , & 74.0 &, \\
\midrule
DDQL            & 1.08 & reference & 0.50 & 110.8 & 11/100 \\
MoE ($K{=}2$)   & 1.06 & $-0.02$ $[-0.15,+0.10]$ & 0.45 & 116.2 & 3/100 \\
MoE ($K{=}4$)   & 1.15 & $+0.07$ $[-0.07,+0.20]$ & 0.52 & 120.6 & 0/100 \\
MoE ($K{=}8$)   & 1.05 & $-0.03$ $[-0.15,+0.09]$ & \textbf{0.37} & 122.2 & 0/100 \\
DDQL-cap4       & 1.06 & $-0.02$ $[-0.14,+0.11]$ & 0.49 & 112.0 & 7/100 \\
DDQL-cap8       & \textbf{0.95} & $-0.13$ $[-0.26,+0.00]$ & 0.50 & \textbf{104.1} & 4/100 \\
\bottomrule
\end{tabular}
\end{table}

\begin{figure}[t]
\centering
\includegraphics[width=\columnwidth]{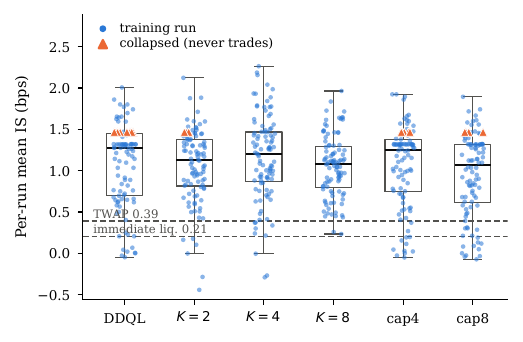}
\Description{Box plot of per-run mean implementation shortfall across 100 training runs for six configurations: DDQL, MoE at K=2, 4 and 8, and capacity-matched dense networks cap4 and cap8. Triangles mark runs whose result vector matches the never-trade policy. Dashed horizontal lines mark the TWAP and immediate-liquidation baselines.}
\caption{Across-seed distribution of per-run mean IS in the main 100-run CUDA campaign ($\phi{=}0.5$). Triangles mark runs whose 500-episode result vector is identical to the never-trade policy's; dashed lines show the TWAP and immediate-liquidation baselines on the same episodes.}
\label{fig:dispersion}
\end{figure}

\textbf{Statistical inference.} No configuration improves mean IS over DDQL: the paired differences and 95\% bootstrap intervals in Table~\ref{tab:ablation} all straddle zero, and the data do not establish equivalence: the $K{=}8$ interval admits $\pm 0.1$ bps, roughly 10\% of the mean. On dispersion, we treat the five variance comparisons against DDQL as one family, with mean and tail comparisons forming separate families; these were defined after the analysis rather than preregistered, so adjusted $p$-values are descriptive rather than confirmatory. Within that family, $K{=}8$ is the only configuration with lower dispersion than DDQL at the unadjusted level (Brown--Forsythe $p{=}0.034$) and does not survive Holm adjustment ($p_{\mathrm{adj}}{=}0.17$); $K{=}4$ vs.\ $K{=}8$, a separate family, gives $p{=}0.0033$.

\textbf{Ties.} Exact duplicate episode vectors occur in 35 DDQL, 24 cap4, 19 cap8 and 3 MoE $K{=}2$ runs, and none in either MoE $K{\geq}4$ arm. These ties do not reduce the 100 seed-level runs, but their discreteness motivates run-level permutation and bootstrap inference for mean and tail comparisons rather than asymptotic tests, and is a further reason to read the Brown--Forsythe dispersion result as exploratory.

The two dispersion columns of Table~\ref{tab:ablation}, shown in Figure~\ref{fig:dispersion}, point in opposite directions: the SD column is variation \emph{across} the 100 trained policies, while CVaR-95 is computed \emph{within} each policy and averaged, asking how bad one trade can get given the policy you hold. MoE $K{=}8$ has the lowest across-seed dispersion of the six and the worst within-policy tail risk; DDQL-cap8 is the reverse. Tail inference uses a paired sign-flip permutation test on the per-run CVaR-95 values over the 100 runs: against DDQL the differences are $+5.4$ bps ($p{=}0.18$), $+9.9$ ($p{=}0.023$) and $+11.5$ ($p{=}0.006$) at $K{=}2,4,8$, versus $+1.3$ ($p{=}0.80$) for cap4 and $-6.6$ ($p{=}0.21$) for cap8. MoE's tail deterioration is monotone in $K$ and significant from $K{=}4$; cap8's advantage is directional only. Pooled percentiles follow the same CVaR ordering at P90, P95 and P99 (cap8 $<$ DDQL $<$ cap4 $<$ $K{=}2 < K{=}4 < K{=}8$), though the pool holds 500 distinct windows, not 50{,}000 independent observations, so we attach no inferential claim. Expert partitioning reduces across-seed dispersion at a real cost in within-policy tails; reporting only run-level SD and calling it ``risk reduction'' overstates what the mechanism delivers.

\textbf{Baseline reference and degenerate runs.} The reference rows of Table~\ref{tab:ablation} record a fact we will not omit: on identical episodes and fill mechanics, TWAP beats every learned configuration on mean IS (0.39 bps) and tail risk (CVaR-95 74.0), and immediate first-step liquidation attains 0.21 bps with a per-run CVaR-95 of 0.8. Two features make early liquidation close to costless: the replay is frictionless, and with the deep-book floor never binding the ten visible levels span a median of only 0.60 bps (Section~\ref{sec:floor}), so executing past the displayed book costs little more than at the touch, while the fifth-best-bid start in the final two steps charges deferred execution an urgency penalty independent of market state. The agents nonetheless fail to find it. The per-step reward is slippage against the \emph{prevailing} mid rather than shortfall against the arrival price, so never trading collects zero reward until one terminal penalty, and the original schedule leaves $\epsilon\geq0.88$ even at the maximum 25{,}000 training actions, far above its intended floor (Table~\ref{tab:hyperparams}). 25 of 600 runs (11 DDQL, 3 MoE $K{=}2$, 7 cap4, 4 cap8, none in either MoE $K{\geq}4$ arm) match this never-trade policy in all 500 evaluation episodes (IS 1.47 bps; Fisher exact $p{=}0.0007$ for DDQL versus each of $K{=}4$ and $K{=}8$). These are run-level counts and do not establish that every expert is inactive. In this GPU campaign, the per-episode IS-cost-matching proxy estimates an inactivity share of 11.1\% for DDQL, 16.8\%, 22.4\% and 24.9\% at $K{=}2,4,8$, and 8.0\% and 4.0\% for cap4 and cap8, and 84 of 100 $K{=}8$ runs contain at least one never-trade episode. The absence of complete-run collapse coexists with episode-level inactivity. Section~\ref{sec:decomp} tests specification changes and their interactions.

\textbf{Inactivity measurement.} The GPU episode rates above use IS-cost matching to the never-trade reference and have not been directly revalidated from actions. This proxy agreed with action-based labels in all 300{,}000 original CPU baseline episodes. In the corrected CPU campaigns it produced three false positives (two DDQL episodes and one cap4 episode), so cost equality is not a universal inactivity criterion. The corrected DDQL result of 0/50{,}000 inactive episodes below is counted directly from executed quantities; the original GPU estimates remain proxy-based.

Section~\ref{sec:oe_moe} rules out inference-time regime conditioning by construction. Dilution of collapsed runs does not account for the dispersion gap either: the eleven collapsed DDQL runs sit at 1.47 bps, only 0.77 across-seed standard deviations above the DDQL mean, and excluding them leaves the remaining 89 runs with a standard deviation of 0.51 against 0.50 for the full set; removing the collapses slightly \emph{increases} dispersion rather than explaining it. Nor is the pattern consistent with simple averaging across $K$ policies, which would predict monotonicity, whereas the observed sequence is 0.50, 0.45, 0.52, 0.37. The lower dispersion at $K{=}8$ is therefore mechanistically unresolved, and since $K{=}4$ is the weakest MoE configuration on both mean and dispersion, a benefit appearing only at $K{=}8$ may reflect selection across a three-point grid. We treat it as exploratory. Per-expert data dilution is a possible explanation for the monotone tail deterioration ($K{=}8$: 312 of 2{,}500 training episodes per expert on average, and roughly 30 for the smallest), but these comparisons do not establish that mechanism.

\textbf{Cluster interpretation.} The $K{=}4$ centroids separate primarily on realised volatility; spread over mid is near-constant across three clusters, and the fourth coordinate is fixed at 1.0, entering K-Means as zero once standardisation adds $10^{-8}$ to each feature's standard deviation. Cluster sizes are highly unequal. In this GPU campaign with the original router, the smallest expert receives on average 1.8\% ($K{=}4$) and 1.1\% ($K{=}8$) of each run's 3{,}000 routing decisions (2{,}500 training and 500 evaluation episodes; the 300 warm-up episodes that fit the router are not routed), and 91 of 100 $K{=}8$ runs have an expert below 2\%. Thin experts see little data, but an expert this small cannot by itself account for never-trading in a quarter of episodes. The partition does not directly observe the routed episode's conditions. In the device-matched CPU replay, excluding each run's first evaluation episode leaves 49{,}900 episodes per $K$. With trained experts and centroids fixed, routes can differ from fixed-horizon market-feature routing only after an early sell-out: at most 2--4\% of these evaluation routes. This bounds direct routing changes, not effects through training.

\label{sec:floor}
\textbf{Floor-sensitivity check.} Section~\ref{sec:moe_model} flags a deep-book fill-price floor $\phi$ (default 50\% of mid) as a potential source of arbitrary tail penalty. We re-ran 30 of the 100 runs at $\phi\in\{30\%,70\%,90\%\}$. Within the sweep the floor has \emph{no effect whatsoever}: the three values produce results identical episode for episode (maximum absolute difference exactly zero across 90{,}000 evaluations), so $\phi$ never binds. The reason is not that the book is deep relative to the order: median visible 10-level depth is 20.9 units against a $q_0{=}20$ order, 46\% of bars display less than the order size, and dumping the remaining inventory into the thinnest bar overflows the book by 14.6 units. Overflow is common; what makes the floor unreachable is how it is \emph{priced}. The ten visible levels span a median of 0.60 bps, so extrapolating off the bottom costs about 0.067 bps per level and leaves the fill at 0.9997 of mid in the worst bar, well above every $\phi$ in the 30--90\% sweep; even $\phi{=}99.9\%$ would not bind. This is a simulator limitation rather than a robustness result: executing beyond the displayed book is very nearly free here, which is part of why immediate liquidation is hard to beat. The device-matched $\phi{=}50\%$ baseline computed for Section~\ref{sec:decomp} confirms this directly rather than by inference: it is bit-identical to the $\phi{=}30\%$ sweep across all 90 (configuration, run) pairs, maximum absolute difference exactly zero, confirming directly that the floor never binds. One disclosure: the sweep does \emph{not} bit-reproduce the original $\phi{=}50\%$ campaign, because it trained on CPU rather than CUDA. Over the 90{,}000 paired evaluations 61\% of values differ and per-run means shift by up to 1.58 bps, so the documented seeds pin down outcomes within a device but not across devices, which we report as a limitation. Campaign-level statistics are unaffected: over the 30 common runs the two agree on mean IS to within 0.07 bps and on the SD of run-level means to within 0.13 bps, with no significant paired mean or variance difference ($p{\geq}0.15$). The floor conclusion rests on the sweep's internal identity across $\phi$, which the device does not affect.

Taken together, once the router lag is corrected and capacity is matched, the mean gap between MoE and DDQL is not significant at any $K$, and neither the dispersion nor the tail property closes the gap to the TWAP reference.

\subsection{Specification Decomposition}
\label{sec:decomp}
Section~\ref{sec:phase2} highlights three specification choices implicated in the failure mode: a reward that differs from the evaluation metric, a slowly decaying exploration schedule leaving $\epsilon\geq0.88$, and asymmetric replay-buffer retention across arms. We retrain DDQL, MoE $K{=}8$ and DDQL-cap8 under each modification alone, under the reward-and-exploration pair, and under all three (Table~\ref{tab:decomp}). Every decomposition arm trains on CPU, so we compare against a CPU-trained baseline rather than Table~\ref{tab:ablation}'s CUDA campaign, removing the device confound by construction rather than by caveat; the two differ only as between-campaign variation would predict (12/100 and 7/100 collapsed runs for DDQL and cap8, against 11/100 and 4/100). All arms share the seed sequence and the identical 500 windows, so comparisons are seed-paired, and collapse counts are cross-checked with McNemar's exact test alongside Fisher's.

The revised decay factor targets 0.01 after 250 decay events, assuming 25{,}000 actions. Early termination reduces that count: the annealing arm's MoE training-transition logs imply 7.6--8.0 steps per episode and a reconstructed final $\epsilon$ of 0.03 on average (0.011--0.075 across runs), so the floor is a nominal target. The dense arms did not log training actions.

\textbf{Annealing suppresses collapse under the as-reported reward.} It removes complete-run collapse from vanilla DDQL, 12/100 to 0/100. Baseline and modified runs share seed identifiers and evaluation windows, so collapse is paired by seed; we report exact McNemar as primary, $p{=}4.9{\times}10^{-4}$ (Fisher $p{=}3.4{\times}10^{-4}$), with perfectly nested pairing, all 12 baseline collapses vanishing and none appearing that were not there before. At the episode level annealing also goes further than partitioning: on this CPU baseline DDQL's never-trade share falls from 12.0\% to none of 50{,}000 episodes, counted from executed quantities, whereas 26.1\% of MoE $K{=}8$ episodes never trade under the original schedule. Expert partitioning is therefore not necessary to suppress this failure under the as-reported reward; its apparent stability benefit can be reproduced by annealing alone. Attributing the collapse to the exploration schedule alone would nonetheless overstate the evidence, since it returns once the reward is aligned. It also reaches significance for the capacity-matched dense network, which the cross-device comparison could not: cap8 falls 7/100 to 0/100 (exact McNemar $p{=}0.016$; Fisher $p{=}0.014$), versus Fisher $p{=}0.12$ on CUDA. On these windows, annealing brings every configuration below TWAP's 0.39 bps, improving on the corresponding baseline by $-0.85$ (DDQL), $-1.07$ ($K{=}8$) and $-0.82$ (cap8) bps, each $p<10^{-4}$ at the permutation test's 50{,}000-draw floor (paired $t$: $p<10^{-23}$). Per-run CVaR-95 is 47.2, 50.3 and 31.0 against TWAP's 74.0, below every configuration in Table~\ref{tab:ablation} (104--122). We attach no execution claim to this: $K{=}8$'s 0.01 bps is execution at essentially the arrival price, reachable only because the replay returns no feedback and intra-episode timing is free. This modified specification is a different diagnostic, not a better testbed.

\textbf{The specification changes interact non-additively.} Each 30-seed arm is compared against the same 30 CPU seeds, where DDQL collapse is 2/30 and cap8's 3/30; the 100-seed rate above is not seed-matched to these. On that basis the reward and buffer modifications alone show no detectable change in DDQL collapse rates (0/30, $p{=}0.49$; 5/30, $p{=}0.42$), and only annealing removes complete-run collapse at full strength. Adding the reward modification to annealing reverses that benefit: 19/30 (Fisher $p{=}6.2{\times}10^{-6}$, McNemar $p{=}1{\times}10^{-4}$, with all but one baseline collapse also present in the arm), rising to 48/100 with the buffer modification added ($p{=}3.2{\times}10^{-8}$ against the device-matched baseline). DDQL-cap8 moves the same way but only directionally (3/30 to 10/30, Fisher $p{=}0.06$, McNemar $p{=}0.09$: its three baseline collapses are absent from the arm's own, so discordant pairs cut both ways). The reward modification alone destabilises cap8 outright (21/30, $p{=}3.4{\times}10^{-6}$). Mean IS agrees: the pair returns DDQL to 1.08 bps, near its as-reported CPU baseline, from 0.05 and 0.28 separately. Changing replay retention alone has no detectable effect on collapse rates, so the reward-plus-annealing arm establishes a reward-by-exploration interaction; interactions involving replay retention are unresolved, two of the three two-change arms not having been run. No MoE $K{=}8$ run has all 500 evaluation episodes inactive under any of the six specifications, but its share of never-trade episodes is 26.1, 1.1, 0.04, 28.7, 3.5 and 5.4\% in the six rows of Table~\ref{tab:decomp}.

\textbf{Two further observations.} $K{=}8$'s dispersion advantage over DDQL, 0.37 against 0.50 as reported, narrows to 0.42 against 0.44 under annealing, where cap8 is lowest at 0.32: the ordering behind the dispersion claim does not survive a change of specification. The annealing effect was not significant at 30 seeds but was at 100; this shows sensitivity of inference to seed count, without identifying a mechanism from the significance threshold crossing.

\begin{table}[t]
\centering
\caption{Specification decomposition: mean IS (bps) and collapsed-run count under each modification, on the same 500 evaluation episodes and seed pairing as Table~\ref{tab:ablation} (TWAP 0.39, immediate liquidation 0.21). Collapse uses the Table~\ref{tab:ablation} definition; evaluation is unchanged, so the never-trade policy scores 1.47 bps in every arm, and means in high-collapse cells average over two attractors. All arms trained on CPU, so the as-reported row is a device-matched CPU baseline; Table~\ref{tab:ablation}'s CUDA campaign gives 1.08/11, 1.05/0 and 0.95/4. Counts are runs: MoE $K{=}8$'s zeros coexist with never-trade episodes (Section~\ref{sec:decomp}).}
\label{tab:decomp}
\scriptsize
\setlength{\tabcolsep}{3.5pt}
\begin{tabular}{l c cc cc cc}
\toprule
 & & \multicolumn{2}{c}{\textbf{DDQL}} & \multicolumn{2}{c}{\textbf{MoE $K{=}8$}} & \multicolumn{2}{c}{\textbf{DDQL-cap8}} \\
\cmidrule(lr){3-4}\cmidrule(lr){5-6}\cmidrule(lr){7-8}
\textbf{Specification} & $n$ & IS & Coll. & IS & Coll. & IS & Coll. \\
\midrule
As reported        & 100 & 1.13 & 12 & 1.08 & 0 &1.01 & 7 \\
$+$ reward         & 30  & 0.05 & 0  & 0.15 & 0 &1.11 & 21 \\
$+$ annealing      & 100 & 0.28 & 0  & \textbf{0.01} & 0 &0.19 & 0 \\
$+$ buffer         & 30  & 1.19 & 5  & 1.07 & 0 &0.90 & 2 \\
$+$ reward \& anneal & 30 & 1.08 & 19 & 0.06 & 0 &1.12 & 10 \\
$+$ all three      & 100 & 0.93 & 48 & 0.15 & 0 &1.11 & 24 \\
\bottomrule
\end{tabular}
\end{table}

\section{Conclusions and Future Work}
\label{sec:conclusion}

We studied a Mixture of Experts (MoE) framework for cryptocurrency order execution, combining a K-Means episode partition with Double Deep Q-Learning experts, evaluated as a training-stability mechanism rather than a mean-cost improvement. Our principal findings are:

\begin{enumerate}
    \item \textbf{Expert partitioning does not improve mean execution cost, and the dispersion effect is unresolved.} Over 100 out-of-sample runs the mean difference between MoE ($K{=}8$) and DDQL is $-0.03$\,bps, 95\% CI $[-0.15,+0.09]$, and the data do not establish equivalence. Across-seed dispersion is lowest at $K{=}8$ (0.37 vs.\ 0.50\,bps, $p{=}0.034$ uncorrected) but fails family-wise adjustment, is non-monotone in $K$, is not explained by collapsed-run dilution, and narrows under the modified specification. Regime conditioning is ruled out by construction, so we report the dispersion result as exploratory.
    \item \textbf{Across-seed stability and within-policy tail risk are separate properties.} Expert tail risk worsens monotonically with $K$ (per-run CVaR-95 $+11.5$\,bps vs.\ DDQL at $K{=}8$, permutation $p{=}0.006$), possibly reflecting per-expert data dilution, whose causal role is not established here, while the capacity-matched control's better tails are directional only ($-6.6$\,bps, $p{=}0.21$).
    \item \textbf{Under the specification as reported, no learned configuration beats the schedule baselines, and the deficit belongs to the specification.} On the same 500 evaluation episodes, TWAP attains 0.39\,bps mean IS and 74.0 CVaR-95, better than every agent, and immediate liquidation 0.21\,bps; 25 of 600 training runs collapse into a policy that never trades before forced terminal liquidation, and the IS-cost-matching proxy in this GPU campaign estimates that per-episode inactivity rises with $K$ from 11.1\% (DDQL) to 24.9\% ($K{=}8$). Annealing alone removes complete-run collapse and brings DDQL to 0.28\,bps, MoE $K{=}8$ to 0.01 and DDQL-cap8 to 0.19, all below TWAP on these windows, with per-run CVaR-95 at 31--50 against TWAP's 74.0. This is not evidence that RL execution works: 0.01\,bps is execution at essentially the arrival price, attainable only because the replay is frictionless. The correct reading remains diagnostic: the architectures reshape the distribution of training outcomes, while the training specification determines whether any of them clears a schedule baseline.
    \item \textbf{The specification changes interact non-additively.} On a device-matched CPU baseline, annealing alone removes complete-run collapse for both vanilla DDQL (0/100 against 12/100, exact McNemar $p{=}4.9{\times}10^{-4}$) and DDQL-cap8 (0/100 against 7/100, exact McNemar $p{=}0.016$), while the reward and buffer modifications alone show no detectable change in collapse rates on their 30-seed comparisons (0/30 and 5/30 against 2/30). Combining annealing with the reward modification restores collapse at 19/30, and all three give 48/100, with DDQL-cap8 moving the same way but only directionally at this seed budget: the reward-plus-annealing arm establishes a reward-by-exploration interaction, while interactions involving replay retention remain unresolved. No MoE $K{=}8$ run has all 500 evaluation episodes inactive under any of them, but up to 28.7\% of its episodes never trade, so the absence of complete-run collapse does not imply episode-level inactivity is prevented.
    \item \textbf{The negative-result checks came back clean, with two disclosures.} A corrected off-by-one routing-lag bug reversed no finding, and the deep-book fill-price floor never binds at this order size, shown directly rather than by inference from the $\phi$ sweep. The disclosures are a cross-device sensitivity in that sweep alone, affecting no headline result, and sensitivity of collapse-suppression inference to the difference between 30 and 100 seeds (Sections~\ref{sec:phase2}, \ref{sec:decomp}).
\end{enumerate}

\subsection{Limitations}
\label{sec:limitations}

First, the historical replay is frictionless: the agent's own trades have no feedback on subsequent prices or book states, each bar's book is a 5-minute mean-aggregated book rather than an instantaneous one, so the realised-volatility feature is computed on time-averaged rather than instantaneous prices, and only the bid side is modelled at execution time. Second, the environment omits the nonlinear and transient impact documented in the microstructure literature \cite{almgren2003, gatheral2010, obizhaeva2013}. Third, the state omits richer features such as multi-level depth imbalance. Fourth, the sample is a single instrument over two weeks, with evaluation on one. Fifth, the model does not represent macroeconomic shocks or flash crashes \cite{gao2024}. Sixth, every comparison rests on a single fixed draw of 500 evaluation windows shared across models, so pooled tail statistics contain 500 distinct market scenarios, not 50{,}000 independent ones. Seventh, the training reward is not the metric reported at evaluation, a mismatch with real consequences; combined with frictionless replay and the terminal rule, it makes the environment's own optimal policy trivial in a way live markets' is not. Eighth, those 500 windows are heterogeneous: TWAP scores $+0.39$ bps over the full set but $-1.66$ bps over the first 120 episodes, so no subset is representative and we report none. Ninth, the decomposition of Section~\ref{sec:decomp} covers three of six configurations, half its arms use 30 rather than 100 seeds, and two of the three two-change arms were not run.

\subsection{Future Work}
The immediate next step is to complete the factorial decomposition of
reward, exploration, and replay retention. More importantly, the
modified specification should be evaluated in an environment where
execution timing and market impact are endogenous: under the present
frictionless replay, the modified agents approach arrival-price
execution, so their improved scores are diagnostic rather than evidence
of a better execution policy. A reactive-agent configuration of ABIDES
\cite{byrd2020abides} could test feedback effects, subject to validation
of its background agents. Historical replay in JAX-LOB
\cite{frey2023jaxlob} alone would not validate strategic responses to
the agent, although its engine can accept strategic-agent messages.
Live-data paper trading could test operations but would not measure
the market impact of the strategy's own orders. In a validated reactive
environment, contemporaneous liquidity
features and time-step-level soft gating \cite{zong2024,puigcerver2024}
could test whether expert routing provides genuine regime conditioning.

\bibliographystyle{ACM-Reference-Format}
\bibliography{references}

\end{document}